\documentclass[aps,showpacs,prl,draft,12pt]{revtex4}

\newcommand{\ba}{\begin{eqnarray}}
\newcommand{\ea}{\end{eqnarray}}
\newcommand{\bmath}{\begin{mathletters}}
\newcommand{\emath}{\end{mathletters}}
\newcommand{\ban}{\begin{eqnarray*}}
\newcommand{\ean}{\end{eqnarray*}}

\begin{document}

\title{Critical-Point Symmetry in a Finite System}

\author{A. Leviatan$^{1}$ and J.N. Ginocchio$^{2}$}

\affiliation{
$^{1}$Racah Institute of Physics, The Hebrew University,
Jerusalem 91904, Israel\\
$^{2}$Theoretical Division, Los Alamos National Laboratory,
Los Alamos, NM, 87545, U.S.A.}

\date{\today}

\begin{abstract}
At a critical point of a second order phase transition the intrinsic
energy surface is flat and
there is no stable minimum value of the deformation. 
However, for a finite system, we show that there is an effective
deformation which can describe the dynamics at the critical point.
This effective deformation is determined by minimizing the energy 
surface after projection onto the appropriate symmetries. 
We derive analytic expressions for  energies and quadrupole rates which
provide good estimates for these observables at the critical point.
\end{abstract}


\maketitle

\newpage

Dynamical systems such as nuclei can undergo phase transitions
associated with a change of shape of their equilibrium configuration.
It has been recognized recently \cite{iac00,iac01} that it is possible
to formulate the concept of a dynamical symmetry (here meaning solvability
in terms of quantum numbers) for systems at the critical point of
such shape phase transitions. The importance of these critical-point
symmetries lies in the fact that they provide a classification of states
and analytic expressions for observables in regions where the structure
changes most rapidly.
For nuclei, two critical-point symmetries, called E(5) \cite{iac00}
(not the Euclidean group in five dimensions),
and X(5) \cite{iac01}, were considered in the geometric framework of
the collective model. This model involves a Bohr Hamiltonian
which describes the dynamics of a macroscopic quadrupole shape
via a differential equation in the intrinsic quadrupole shape variables
$\beta$ and $\gamma$.
The E(5) [X(5)] critical-point symmetry (CPS)
is applicable to a second- [first-] order shape phase transition between
spherical and deformed $\gamma$-unstable
[axially-symmetric] nuclei.
In the present work we focus on the E(5) CPS
for which an empirical example
has been found in $^{134}$Ba \cite{cas00,arias01} and possibly in
$^{104}$Ru \cite{frank02}, $^{102}$Pd \cite{zamfir02} and
$^{108}$Pd \cite{zhang02}.
To apply the E(5) CPS to real nuclei one has to take into account
the finite number of nucleons. This can be conveniently done in the
algebraic framework of the interacting boson model (IBM) \cite{ibm}.
This model describes low-lying quadrupole collective states in nuclei
in terms of a system of $N$ monopole ($s$) and
quadrupole ($d$) bosons representing valence nucleon pairs.
The IBM Hamiltonian relevant to the critical point of the phase transition
between spherical and $\gamma$-unstable deformed nuclei preserves the
$O(5)$ symmetry
\cite{diep80}.  Its energy surface, obtained by the method of coherent
states
\cite{diep80,gino80}, is $\gamma$-independent and exhibits a flat-bottomed
behavior in $\beta$ which resembles the infinite square-well potential
used to derive the E(5) CPS in the geometric approach
\cite{iac00}.
Calculations with finite $N$ values ($N$=5 for $^{134}$Ba) have found
that this critical IBM Hamiltonian
can replicate numerically the E(5) CPS and
its analytic predictions.
\cite{cas00,frank02,zamfir02,zhang02}.
In the present work we examine
the properties and conditions
that enable features of E(5) CPS to occur in a finite system. For
that purpose we propose wave functions of a particular analytic form,
which can simulate accurately the exact IBM eigenstates at the critical
point. These wave functions with fixed $N$ and good $O(5)$ symmetry
are used to derive accurate estimates for energies and quadrupole rates
at the critical point without invoking large-$N$ approximations.
The proposed wave functions can be obtained by projection from
intrinsic states with an effective $\beta$-deformation.

In the geometric approach
the E(5) eigenfunctions \cite{iac00}
are proportional to
Bessel functions of order $\tau + {3\over 2}$
and the corresponding eigenvalues are proportional to $(x_{\xi,\tau})^2$.
Here $\tau$ is the $O(5)$ quantum number, and
$x_{\xi,\tau}$ is the $\xi$-th root of these Bessel functions.
A portion of an E(5)-like spectrum is shown in Fig.~1. It consists of
states, $L^{+}_{\xi,\tau}$, arranged in major families labeled by
$\xi=1,2,\ldots$ and $O(5)$ $\tau$-multiplets ($\tau=0,1,\ldots$)
within each family. The angular momenta $L$ for each $\tau$-multiplet
are obtained by the usual $O(5)\supset O(3)$ reduction \cite{ibmo6}.
The E(5) CPS leads to analytic parameter-free
predictions for energy ratios and $B(E2)$ ratios
which persist when carried over to a finite-depth
potential \cite{caprio02}. As seen in Table I, the E(5) predicted values
are in-between the values expected of a spherical vibrator [$U(5)$]
and a deformed $\gamma$-unstable rotor [$O(6)$].

In the algebraic approach, the $U(5)$-$O(6)$ transition region is modeled by
the Hamiltonian
\ba
H &=& \epsilon\,\hat{n}_d + {1\over 4}A\,
\left[\, d^{\dagger}\cdot d^{\dagger} -  (s^{\dagger})^2\,\right ]
\left[\, H.c.\,\right]
\label{hamilt}
\ea
with $\epsilon$ and $A$ positive parameters. Here $\hat{n}_d$ is the
$d$-boson number operator, $H.c.$ stands for Hermitian conjugate and
the dot implies a scalar product.
In the $U(5)$ limit ($A=0$), the spectrum of $H$ is harmonic,
$\epsilon\, n_d$, with $n_d=0,1,2\ldots N$. The eigenstates are
classified according to the chain
$U(6)\supset U(5)\supset O(5) \supset O(3)$ with quantum numbers
$\vert\,N,n_d,\tau,L\rangle$ (for $\tau\geq 6$ an additional multiplicity
index is required for complete classification).
These states can be organized into sets characterized by $n_d=\tau +2k$.
States in the lowest-energy set ($k=0$) satisfy
$P_{0}\,\vert\,N,n_d=\tau,\tau,L\rangle = 0$ with
$P_{0}^{\dagger}=d^{\dagger}\cdot d^{\dagger}$.
Other sets ($k >0$) are generated
by $\vert\,N,n_d,\tau,L\rangle \propto
(P^{\dagger}_{0})^{k}\vert\,N-2k,n_d=\tau,\tau,L\rangle$.
In the $O(6)$ limit ($\epsilon=0$), the spectrum is
${1\over 4}A(N-\sigma)(N+\sigma+4)$ with $\sigma=N,N-2,N-4,\ldots 0\;
{\rm or}\; 1$. The eigenstates are classified according to the chain
$U(6)\supset O(6)\supset O(5) \supset O(3)$ with quantum numbers
$\vert\,N,\sigma,\tau,L\rangle$. The ground band has $\sigma = N$ and
its members satisfy
$P_{1}\,\vert\,N,\sigma=N,\tau,L\rangle = 0$ with
$P^{\dagger}_{1} =
[\, d^{\dagger}\cdot d^{\dagger} - \, (s^{\dagger})^2\,]$.
The remaining bands with $\sigma=N-2k$ are generated by
$\vert\,N,\sigma,\tau,L\rangle
\propto (P^{\dagger}_{1})^{k}\,\vert\,N-2k,\sigma ,\tau,L\rangle.$
These results suggest that in-between the $U(5)$ and $O(6)$ limits,
we consider a ground band ($\xi=1$) for the Hamiltonian ({\ref{hamilt}})
determined by the condition
\ba
&&P_{y}\,\vert\, \xi=1;y,N,\tau,L\rangle = 0 ~,
\nonumber\\
&&P^{\dagger}_{y} =
\left[\, d^{\dagger}\cdot d^{\dagger} - y\, (s^{\dagger})^2\,\right ].
\label{p0}
\ea
In the $U(5)$ basis these states are
\ba
\vert\, \xi=1;y,N,\tau, L\rangle &=&
\sum_{n_d}{1\over 2}\left [1 + (-1)^{n_d-\tau}\right ]\, \xi_{n_d,\tau}
\vert\, N,n_d,\tau, L\rangle ~,
\label{projt}
\ea
and the $n_d$ summation covers the range $\tau\leq n_d\leq N$.
The coefficients $\xi_{n_d,\tau}$ have the explicit form
\ba
\xi_{n_d,\tau} &=& \left [ (N-\tau)!\,(2\tau+3)!!\over
(N-n_d)!\,(n_d-\tau)!!\,(n_d+\tau+3)!!\,\right ]^{1/2}\,
y^{(n_d-\tau)/2}\,\xi_{\tau,\tau} ~,
\nonumber\\
\left (\xi_{\tau,\tau}\right )^2 &=& {2(N-\tau+1)\over (2\tau+3)!!}\,
y^{2\tau+3}\,\left [ G^{(\tau+1)}_{N+1-\tau}(y)\right ]^{-1} ~ ,
\nonumber\\
G^{(n)}_{\alpha}(y) &=&2 y^{2n+1}\sum_{p}\left({\alpha\atop 2p+1}\right )
\, y^{2p}\, {(2p+1)!!\over (2p+ 2n +1)!!} ~.
\label{xindt}
\ea
Members of the first excited band $(\xi=2)$ have approximate
wave functions of the form
\ba
&&\vert\, \xi=2; y, N,\tau, L\rangle = 
{\cal N}_{\beta}\,P^{\dagger}_{y}\,\vert\, \xi=1; y,N-2,\tau,L\rangle
\nonumber\\
&&{\cal N}_{\beta} = \left [ \, 2(2N+y^2 +1) +
4(y^2-1)S^{(N-2)}_{1,\tau}\, \right ] ^{-1/2} ~,
\label{projtbeta}
\ea
where $S^{(N)}_{1,\tau}$ is defined in Eq.~(\ref{s1k}) below.

The states of Eqs.~(\ref{projt}) and (\ref{projtbeta}) have fixed $N$,
$L$ and good $O(5)$ symmetry $\tau$.
Henceforth, for reasons to be explained below,
they will be referred to as
$\tau$-projected states. Diagonal matrix elements of the Hamiltonian
(\ref{hamilt}) in these states, denoted by $E_{\xi,\tau}=
\langle \xi; y, N, \tau\vert H \vert \xi; y, N, \tau\rangle $,
can be evaluated in closed form
\ba
E_{\xi=1,\tau}&=&
\epsilon\left[\, N - S^{(N)}_{1,\tau}\,\right ]
+ {1\over 4}A\, (1-y)^2\, S^{(N)}_{2,\tau}
\nonumber\\
E_{\xi=2,\tau}&=&
\epsilon\left\{ N - 2{\cal N}^{2}_{\beta}\,
\left [\, 2y^2 + (2N + 7y^2 -1)\,S^{(N-2)}_{1,\tau}
+ 2(y^2-1)\,S^{(N-2)}_{2,\tau}\,\right ]\,\right\}
\nonumber\\
&&\;\;\; +{1\over 4}A\,2{\cal N}^{2}_{\beta}\,
\left\{\, 2(y-1)^2\,(y^2-1)\,S^{(N-2)}_{3,\tau}
+ (y-1)^2\,(2N+y^2 -8y + 5)\,S^{(N-2)}_{2,\tau}\right.
\nonumber\\
&&\;\;\; \left. \qquad\qquad\quad
+ 16(y-1)(N+y)\,S^{(N-2)}_{1,\tau}
+ 2\left [\,(2N+y)(2N+y+2) + 1\,\right ]\right\} ~.
\label{enexit}
\ea
The $E_{\xi,\tau}$
are independent of $L$ since $H$ is an $O(5)$ scalar. Their
expressions involve the quantities
$S^{(N)}_{k,\tau} =\langle \xi=1; y,N,\tau\vert (s^{\dagger})^ks^k
\vert \xi=1;y, N, \tau\rangle$ which are given by
\ba
S^{(N)}_{1,\tau} &=&
(N-\tau + 1) {G^{(\tau+1)}_{N-\tau}(y)\over
G^{(\tau+1)}_{N-\tau+1}(y)}
\label{s1k}
\ea
with $S^{(N)}_{2,\tau} = S^{(N)}_{1,\tau}S^{(N-1)}_{1,\tau}$ and
$S^{(N)}_{3,\tau} =
S^{(N)}_{1,\tau}S^{(N-1)}_{1,\tau}S^{(N-2)}_{1,\tau}$.
Non-diagonal matrix elements of $H$ between $\tau$-projected states
in different $\xi$-bands, $H_{1,2;\tau}=
\langle \xi=2; y, N, \tau\vert H \vert \xi=1; y, N, \tau\rangle$,
can be evaluated as well
\ba
H_{1,2;\tau} &=&
2{\cal N}_{\beta}\left[S^{(N)}_{2,\tau}\right]^{1/2}
\left\{\, \epsilon\,y + {1\over 4}A(y-1)\left[\, (2N+y+1)
+ 2(y-1)\, S^{(N-2)}_{1,\tau}\,\right]\right\} ~.
\label{nondiag}
\ea
By techniques similar to that employed in the $O(6)$ limit of the IBM
\cite{ibmo6}, explicit expressions of quadrupole rates can be derived for
transitions between the $\tau$-projected states. For the relevant IBM
quadrupole operator, $T(E2) = d^{\dagger}s + s^{\dagger}\tilde{d}$, these
transitions are subject to the $O(5)$ selection rule $\Delta\tau=\pm 1$,
and, as explained in the caption of Fig.~1, it is sufficient to focus
on the B(E2) values of the type
\ba
&&B(E2;\, \xi=1;\, \tau+1, L=2\tau+2
\to \xi=1,\, \tau,L=2\tau) =
\nonumber\\
&&\qquad\qquad\qquad
{(\tau+1)\over (2\tau+5)(N-\tau+1)}\,
\left ( S^{(N)}_{1,\tau}\right )^2\,
{G^{(\tau+1)}_{N-\tau+1}(y)\over G^{(\tau+2)}_{N-\tau}(y)}
\left [ \, y + (N-\tau)
{G^{(\tau+2)}_{N-\tau-1}(y)\over G^{(\tau+1)}_{N-\tau}(y)}\right ]^2 ~,
\nonumber\\
&&B(E2;\, \xi=2,\, \tau, L=2\tau\to 
\xi=1,\, \tau+1,L=2\tau+2) =
\nonumber\\
&&\qquad\qquad\qquad
{(\tau+1)(4\tau+5)\over (4\tau+1)(2\tau+5)}
4\,{\cal N}^{2}_{\beta}\, y^2\,(y-1)^2\, (N-\tau)\,
{G^{(\tau+1)}_{N-\tau-1}(y)\over G^{(\tau+2)}_{N-\tau}(y)} ~.
\label{be2xit}
\ea

The states in Eq.~(\ref{projt})
and Eq.~(\ref{projtbeta}) can be obtained by $O(5)$
projection from the IBM intrinsic states for the ground band
\ba
\vert\,c; N \rangle &=&
(N!)^{-1/2}(b^{\dagger}_{c})^N\,\vert 0\,\rangle
\nonumber\\
b^{\dagger}_{c} &=& (1+\beta^2)^{-1/2}\left [\,
\beta\,\cos\gamma\, d^{\dagger}_{0} + \beta\,\sin{\gamma}\,
{1\over\sqrt{2}}\left ( d^{\dagger}_{2} + d^{\dagger}_{-2}\right )
    + s^{\dagger}\,\right]
\label{cond}
\ea
and respectively for the $\beta$ band
\ba
\vert\,\beta; N \rangle &=&
{\cal N}_{\beta}P^{\dagger}_{y}\vert\, c; N-2\rangle ~,
\label{betaint}
\ea
provided $y=\beta^2$.
The expressions in Eqs.~(3)-(9)
depend on the so far unspecified parameter $y$.
Normally, the equilibrium value of $\beta$, and hence $y$, is chosen as
the global minimum of the intrinsic energy surface 
determined from the expectation value of $H$ 
in the intrinsic state (\ref{cond}).
This is a standard procedure for a Hamiltonian
describing nuclei with rigid shapes,
for which the global minimum is deep and well-localized. However,
near the critical point of the phase transition
an alternative procedure is required.

The IBM Hamiltonian, $H_{cri}$, at the critical point of the
$U(5)$-$O(6)$ phase transition corresponds \cite{diep80} to a special
choice of parameters in the Hamiltonian of Eq.~(\ref{hamilt})
\ba
H_{cri}:\quad \epsilon = (N-1)A ~.
\label{hcri}
\ea
The intrinsic energy surface of $H_{cri}$ has the form
\ba
E(\beta) &=& E_0 + A\,N(N-1)\,\beta^4(1+\beta^2)^{-2}
\label{enesurf}
\ea
with $E_0 = \frac{1}{4}AN(N-1)$ a constant. The energy
surface $E(\beta)$, shown in Fig.~(2a), has a flat
behavior ($\sim \beta^4$) for small $\beta$, an inflection
point at $\beta=1$ and approaches a constant for large $\beta$.
The global minimum at $\beta=0$ is not well-localized
and $E(\beta)$ exhibits considerable instability in $\beta$,
resembling a square-well potential for $0\leq\beta\leq 1$.
Under such circumstances fluctuations in $\beta$ are large and
play a significant role in the dynamics. Some of their effect can
be taken into account by introducing into the intrinsic states of
Eqs.~(\ref{cond}) and (\ref{betaint}) an effective $\beta$-deformation.
The effective deformation is expected to be in the range
$0 < y=\beta^2 < 1$, in-between the respective $U(5)$ and $O(6)$ 
value of $\beta$. This will enable a reproduction of E(5) characteristic
signatures which are in-between these limits (see Table I).
In contrast to $E(\beta)$, we see from Fig.~(2b) that the
$O(5)$ projected energy surface $E_{\xi=1,\tau=0}(y)$ of $H_{cri}$
(Eqs.~(\ref{enexit}) and (\ref{hcri}) with $N=5$),
does have a stable minimum at a certain value of $y$,
which we interpret as an effective $\beta$-deformation.
This procedure, based on variation after projection, is in the spirit
of \cite{otsuka87} in which it is shown
that in finite boson systems, a $\gamma$-unstable $O(6)$ state can be
generated from a rigid triaxial intrinsic state with an effective
$\gamma$-deformation of 30$^{\circ}$.
In the present case the $\gamma$-instability is treated exactly by means
of $O(5)$ symmetry, while the $\beta$-instability is treated by
means of an effective deformation.
The appropriate value of $y$ can be used to evaluate the band-mixing,
$\eta_{\tau}(y)={|H_{1,2;\tau}|\over E_{2,\tau}- E_{1,\tau}}$.
A small value of $\eta_{\tau}$ will ensure that the
$\tau$-projected states of Eqs.~(\ref{projt}),
(\ref{projtbeta}) form a good representation of the actual eigenstates of
$H_{cri}$, and turn the expressions of Eqs.~(\ref{enexit}), (\ref{be2xit})
into meaningful estimates for energies and quadrupole transition rates at
the critical point.

To test the suggested procedure we compare in Table II the $U(5)$
decomposition of exact eigenstates obtained from numerical diagonalization
of $H_{cri}$ for $N=5$ with that calculated from the $\tau$-projected states
with $y=0.314$ [the global minimum of $E_{1,0}(y)$].
As can be seen, the latter provide a good approximation to
the exact eigenstates (the corresponding band-mixing
is $\eta_{\tau}=0.12,\,3.53,\,4.14,\,3.05\%$ for
$\tau=0,1,2,3$).
This agreement in the structure of wave functions is
translated also
into an agreement in energies and B(E2) values as
shown in Table I.
The results of Table I and II clearly demonstrate
the ability
of the suggested procedure to provide analytic and
accurate estimates
to the exact finite-N calculations of the critical
IBM Hamiltonian, which
in-turn agree with the experimental data in
$^{134}$Ba and captures the
essential features of the E(5)
critical-point symmetry.

To summarize, in this work we have
considered properties of a critical-point symmetry in a finite
system. We have focused on the E(5) critical-point symmetry
corresponding to flat-bottomed potentials as encountered in a
second-order phase transition between spherical and deformed
$\gamma$-unstable nuclei. We have shown that intrinsic states 
with an effective $\beta$-deformation, reproduce the dynamics 
of the underlying non-rigid shapes. The effective deformation 
can be determined from the global minimum of the energy surface 
after projection onto the appropriate symmetry. 
In the present case, states of fixed N and good $O(5)$ symmetry 
projected from these intrinsic states provide good analytic 
estimates to the exact eigenstates, energies and quadrupole 
transition rates at the critical point.

This work was supported in part by the Israel Science Foundation 
(A.L.) and in part by the U.S. Department of Energy under
contract W-7405-ENG-36 (J.N.G).

\begin{table}
\caption{
Excitation energies (normalized to the energy of the
first excited state) and B(E2) values (in units of
$B(E2;\, 2^{+}_{1,1}\to 0^{+}_{1,0})=1$)
for the E(5) critical-point symmetry {\protect\cite{iac00}}, 
for several N=5 calculations and for the experimental data of 
$^{134}$Ba {\protect\cite{ba134}}. The finite-N calculations involve 
the exact diagonalization of the critical 
IBM Hamiltonian ($H_{cri}$) [Eq.~(\ref{hcri})], $\tau$-projected states 
for $H_{cri}$ [Eqs.~(\ref{enexit}),(\ref{be2xit}) with $y=0.314$], 
the $U(5)$ limit [$\epsilon\,n_d$] and the $O(6)$ limit 
[$(A/4)(N-\sigma)(N+\sigma+4)+B\tau(\tau+3)$].
\normalsize}
\vskip 10pt
\begin{ruledtabular}
\begin{tabular}{lcccccc}

& E(5) & exact &
$\tau$-projection & $U(5)$ & $O(6)$ & $^{134}$Ba \\
&  & N=5 & N=5 & N=5  & N=5  & exp \\
\hline
$E(0^{+}_{1,0})$ & 0    & 0     & 0     & 0  & 0   & 0    \\
$E(2^{+}_{1,1})$ & 1    & 1     & 1     & 1  & 1   & 1    \\
$E(L^{+}_{1,2})$ & 2.20 & 2.195 & 2.19  & 2  & 2.5 & 2.32 \\
$E(L^{+}_{1,3})$ & 3.59 & 3.55  & 3.535 & 3  & 4.5 & 3.66 \\
$E(0^{+}_{2,0})$ & 3.03 & 3.68  & 3.71  & 2  & 
$1.5{A\over B}$ & 3.57  \\
\hline
$B(E2;\, 4^{+}_{1,2}\to 2^{+}_{1,1})$ &
1.68 & 1.38 & 1.35 & 1.6 & 1.27 & 1.56(18) \\
$B(E2;\, 6^{+}_{1,3}\to 4^{+}_{1,2})$  & 
2.21 & 1.40 & 1.38 & 1.8 & 1.22 &  \\
$B(E2;\, 0^{+}_{2,0}\to 2^{+}_{1,1})$ & 
0.86 & 0.51 & 0.43 & 1.6 & 0 & 0.42(12) \\
\end{tabular}
\end{ruledtabular}
\end{table}

\clearpage

\begin{table}
\caption[]{$U(5)$ decomposition (in \%) of the $L^{+}_{\xi,\tau}$ 
states for $N=5$. 
The calculated values are obtained from the $\tau$-projected states, 
Eqs.~(\ref{xindt}), (\ref{projtbeta}) with $y=0.314$. 
The exact values are obtained from numerical diagonalization of the 
critical IBM Hamiltonian $H_{cri}$, Eq.~(\ref{hcri}).
\normalsize}
\vskip 10pt
\begin{ruledtabular}
\begin{tabular}{rcccccccccc}
&\multicolumn{2}{c}{$0^{+}_{1,0}$}
&
\multicolumn{2}{c}{$2^{+}_{1,1}$}
&
\multicolumn{2}{c}{$L^{+}_{1,2}$}
&
\multicolumn{2}{c}{$L^{+}_{1,3}$}
&
\multicolumn{2}{c}{$0^{+}_{2,0}$}
\\
\cline{2-3}\cline{4-5}\cline{6-7}\cline{8-9}\cline{10-11}
$n_d$
&
calc & exact & calc & exact &
calc & exact &
calc
&
exact & calc & exact \\
\hline
0 & 83.2 & 83.4 &      & &
&      &      &      & 15.8 & 16.4 \\
1&      &      & 92.2 & 90.2 &
&      &      &      &      &      \\
2 & 16.4 & 16.2 &
&      & 96.8
& 95.2 &      &      & 70.9 & 76.2 \\
3 &      &      &
7.8  & 9.7  &
&      & 99.1 & 98.4 &      &      \\
4 &  0.4 &
0.4 &      &      &
3.2  & 4.8  &      &      & 13.3 &  7.4 \\
5 &      &      & 0.0  & 0.1  &
&      &  0.9 &  1.6 & &
\end{tabular}
\end{ruledtabular}
\end{table}

\clearpage

\begin{figure}
\caption{An 
E(5)-like spectrum of states labeled by $L_{\xi,\tau}$. 
Shown are the transitions whose B(E2) values are
given in Eq.~(\ref{be2xit}).
The E2 rates for other $\Delta\tau=1$
transitions (not shown) are governed by $O(5)$ symmetry.
Specifically, $B(E2;\, L^{+}_{1,2}\to 2^{+}_{1,1})$
for $L=4,2$ are in the ratio $1:1$ respectively,
$B(E2;\, L^{+}_{1,3}\to 4^{+}_{1,2})$ for $L=6,4,3$
and $B(E2;\, L^{+}_{1,3}\to 2^{+}_{1,2})$ for $L=4,3,0$
are in the ratios $1: 10/21: 2/7: 11/21: 5/7: 1$,
respectively.}
\end{figure}

\begin{figure}
\caption{
Energy
surfaces
of the critical IBM Hamiltonian $H_{cri}$
(\ref{hcri}) with
$N=5$ and
$A=1$.
(a)~Intrinsic energy surface
$E(\beta)$,
Eq.~(\ref{enesurf})
[solid line], and its approximation
by a
square-well
potential [dashed line].
(b)~$O(5)$ projected
energy
surface $E_{\xi=1,\tau=0}(y)$,
Eq.~(\ref{enexit}). The global
minimum
is at $y=0.314$.}
\end{figure}

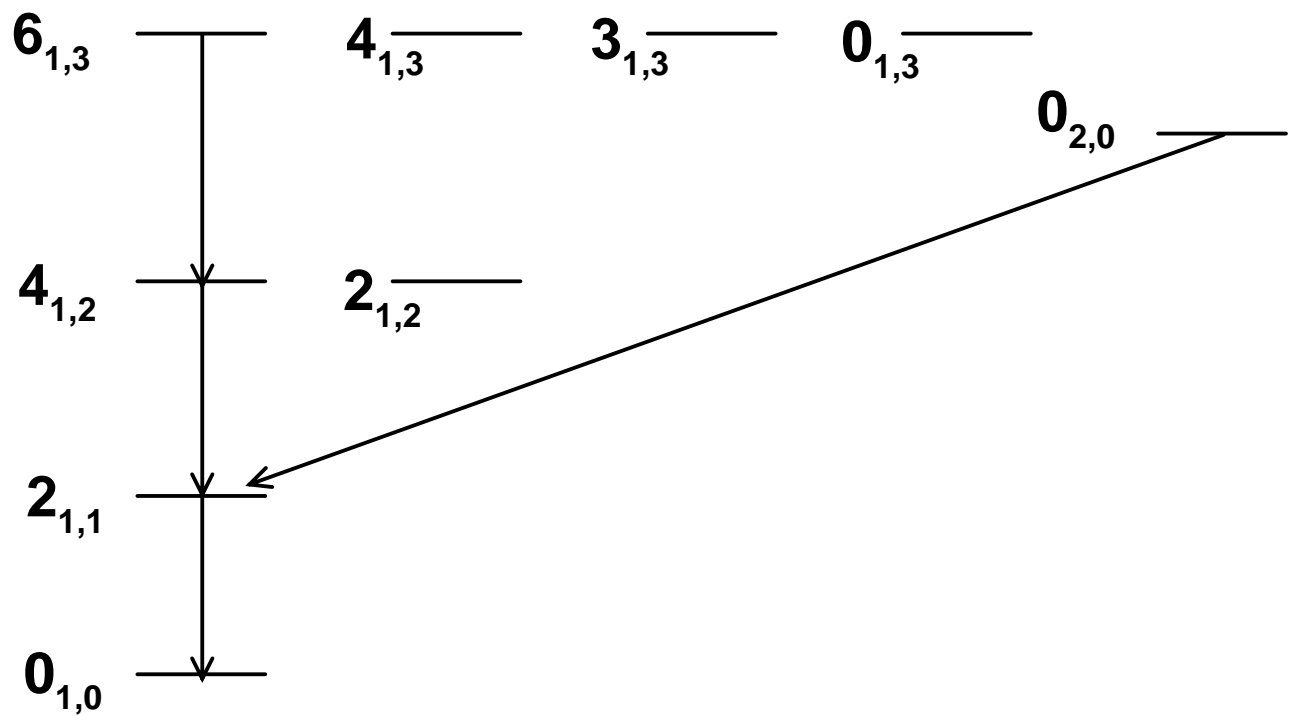
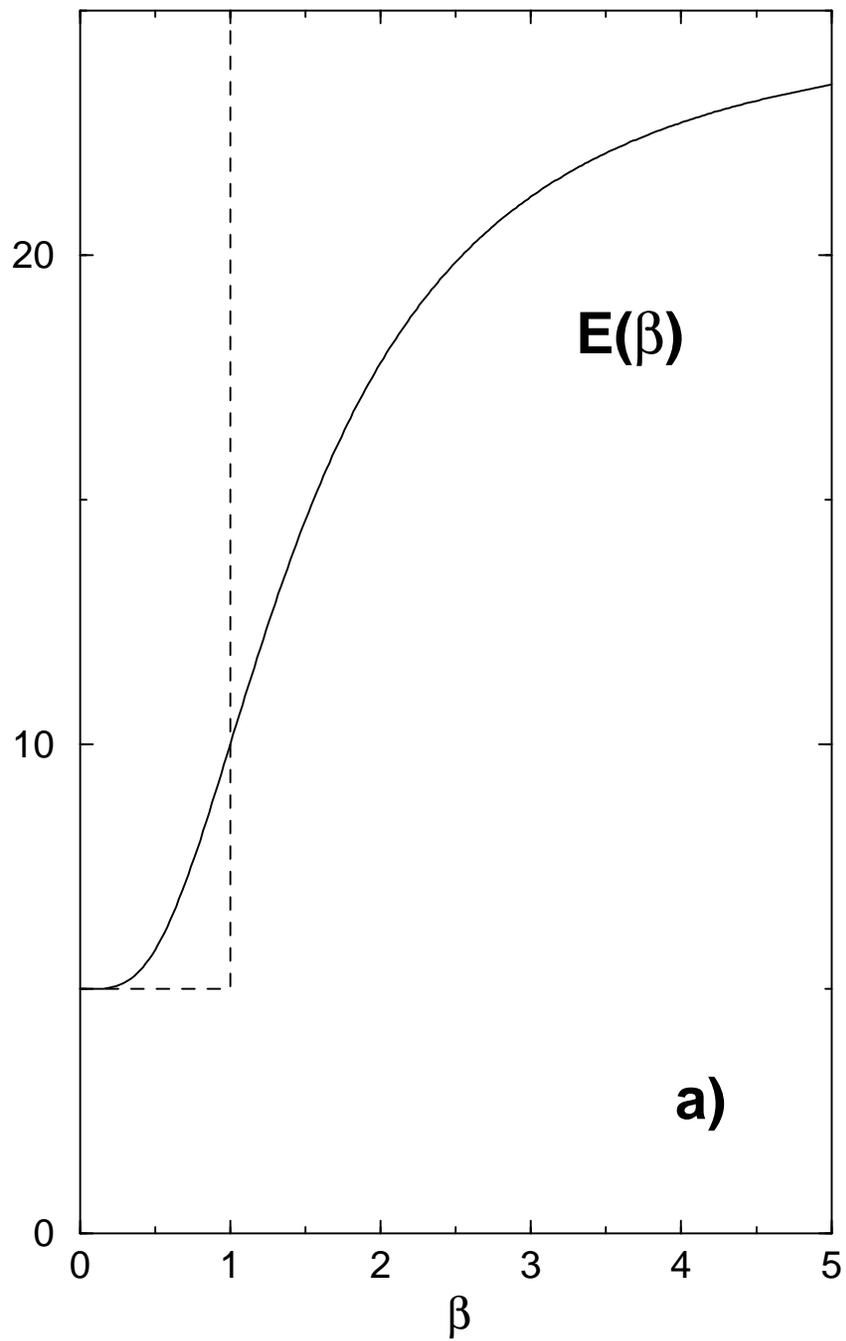
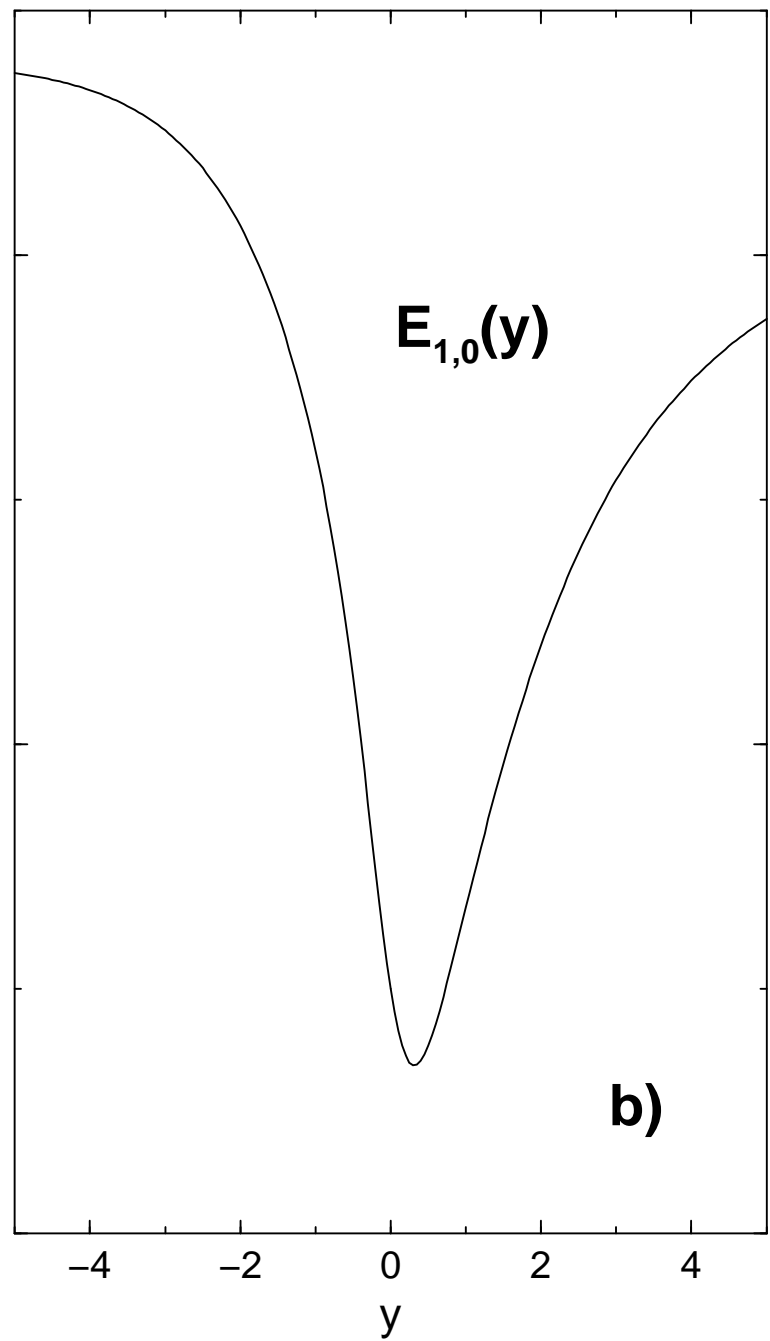
\end{document}